# Gravitational Wave Detection and Fundamental Physics in Space: FOREWORD

Bala Iyer (Raman Research Institute) and Wei-Tou Ni (National Tsing Hua University)

Ever since the completion of special relativity, gravitational waves (GWs) are mentioned. Shortly after the final formulation of general relativity, Einstein predicted the existence of gravitational waves and estimated its strength in his two seminal papers in 1916 and 1918, although he did not think it was experimentally detectable due to its strength compared to sensitivity of instruments available. It is almost one hundred years now and we have seen that a flourish of GW detection activities goes hand in hand with technological advances and astrophysical understandings. The effect on the orbit shrinkage due to GW emission has been first observed in the binary pulsar PSR 1913+16.[1] Efforts for direct detection have been put forward into ground experiments, space missions and Pulsar Timing Arrays (PTAs). The endeavor to detect GWs by terrestrial detectors has been ongoing for more than 50 years and the GW community is on the verge of a direct detection, likely in between 2016-2018 to celebrate the centennial prediction of GWs in general relativity.

Ground based GW detectors probe high-frequency GWs. Since gravitation is the dominant force shaping the large scale structure and the evolution of cosmos, the detection of GWs generated in these processes is needed to probe the large scale or earlier epoch phenomena in the cosmos. Space provides a larger arena and quieter environment for having longer arms and larger detectors for this purpose to probe the low-frequency GWs. The first public proposal on space interferometers for GW detection was presented in the Second International Conference on Precision Measurement and Fundamental Constants (PMFC-II), 8-12 June 1981 in Gaithersburg.[2,3] In this seminal proposal, Faller and Bender studied possible gravitational wave mission concepts in space using laser interferometry. Two basic ingredients were addressed – drag-free navigation for reduction of perturbing forces on the spacecraft (S/C) and laser interferometry for sensitivity of measurement. The LISA-like S/C orbit formation (LISA: Laser Interferometric Space Antenna)[4] was reached in 1985 for the proposal LAGOS (Laser Antenna for Gravitational-radiation Observation in Space).[5] It is natural for people working in lunar laser ranging and measuring free fall acceleration using interferometry to propose such an experiment. In fact, test mass free fall inside a falling shroud in vacuum in the interferometric measurement of the earth gravitational acceleration can be considered as a passive drag-free navigation device. The discrepancy in the absolute gravimeter comparison at BIPM (Bureau International des Poids et Mesures) is partially resolved using correction to interferometric measurements of absolute gravity arising from the finite speed of light.[6] In the S/C tracking, finite velocity of light has always been incorporated. Both test mass for GW missions and test mass of interferometric gravimeter can be regarded as freely falling objects in the solar system and tracked using astrodynamical equation. Thus, we see the interplay among EP (Equivalence Principle) experiments, space geodesy and GW detection missions.

A big step for the GW detection in space is the 1993 ESA M3 Assessment study of LISA and later recommendation as the third cornerstone of "Horizon 2000 Plus". After 2000, LISA became a joint ESA-NASA mission until the 2011 NASA withdrawal. In 1998, LISA Pathfinder was selected as the second of the European Space Agency's Small Missions for Advanced Research in Technology (SMART) to develop and test the demanding drag-free technology. At the Centennial Celebration of General Relativity in 2015, we look forward to the launch of LISA Pathfinder to prepare the road for future spaceborne GW detectors. Based on the ongoing technological development for LISA Pathfinder, ESA has sponsored a technology reference study for the fundamental physics explorer as a common bus for fundamental physics missions.[7] NGO/eLISA (NGO: New Gravitational-wave Observatory), down-scaled from 5 million km arm length to 1 million km arm length, was proposed in 2011 to accommodate the budget change and received excellent evaluation. Most people believe that in time NGO/eLISA will be the first GW mission to fly.

The general concept of ASTROD (Astrodynamical Space Test of Relativity using Optical Devices) is to have a constellation of drag-free S/C navigate through the solar system and range with one another using optical devices to map the solar-system gravitational field, to measure related solar-system parameters, to test relativistic gravity, to observe solar g-mode oscillations, and to detect gravitational waves. A baseline implementation of ASTROD was proposed in 1993 and has been under concept and laboratory studies since then. In 1996, ASTROD I (Mini-ASTROD) with one S/C ranging with ground stations was proposed for testing relativistic gravity and mapping the solar system. The mission study shows that precision of testing relativistic gravity in the solar-system is achievable to $10^{-8}$-$10^{-9}$ in term of Eddington parameter $\gamma$, a more than three orders of improvement over the present precision, and in other aspects with accompanying improvement. In 2009, a dedicated mission concept ASTROD-GW for GW detection with nominal arm length of 260 million kilometers was proposed.

DECIGO (DECi-hertz Interferometer Gravitational-wave Observatory) was proposed in 2001 with the aim of detecting GWs from the early universe and in the observation band between those of terrestrial band and other space GW detectors. It will use a Fabry-Perot method (instead of a delay line method) as in the ground interferometers but with a 1000 km arm length. As a LISA follow-on BBO (Big Bang Observer) was proposed in US with a similar goal. A likely version of DECIGO/BBO is to have 12 S/C with correlated detection. 6S/C-ASTROD-GW has also been considered to possibly explore the relic GWs in the lower part of the low frequency band.[8]

In this special issue, we start with three overview papers on LISA Pathfinder, DECIGO Pathfinder and ASTROD I followed by an overview paper on ASTROD-GW and a paper on the deployment of S/C formation to 1 AU orbit for ASTROD-GW. The deployment issue is a common one for DECIGO, BBO and ASTROD-GW. In the discussion of this issue, it is clear that the propellant ratio can be optimized to about 0.5 to reach anywhere in the 1 AU orbit within a reasonable time of about 1.5 years. The fuel requirement is significant but manageable. NGO and all the above mission proposals use drag-free navigations. Particle detectors are proposed to monitor the charging process of the freely floating proof-masses. These particle detectors will naturally provide SEP (Solar Energetic Particle) observations at different heliocentric longitude and distances from Earth for space weather applications. A repertoire paper addresses the radiation monitor payload and its application to solar physics and space weather.

It is remarkable that the conception of using pulsar timing for the detection of GWs and for testing the foundation of relativistic gravity occurred at about the same period of time as the conception of space GW proposals.[9] They all occurred shortly after the method of using Doppler tracking of S/C to constrain the amplitude of GWs. Here is another example of connection between the testing fundamental physics and the detecting GWs. A big step in using pulsar timing arrays for detecting gravitational waves came when the millisecond pulsars were discovered. Millisecond pulsars usually emit pulses with ultra-stable stability and serve as good clocks. In this special issue, we have two repertoire papers on this subject --- one on pulsar searching and timing, and one on pulsar timing arrays. The present upper limits on the background from PTAs reach $10^{-15}$ level of the characteristic strain in the very low frequency band and already rule out models in which giant elliptical galaxies grow by merger alone. As the observations accumulate, PTAs have the chance to detect the GW background from the Massive Black Hole (MBH) coevolution with Galaxies anytime.[10]

ASTROSAT is an astronomy satellite designed for simultaneous multi-wavelength studies in the Optical/UV and a broad X-ray energy range. With four X-ray instruments and a pair of UV-Optical telescopes, ASTROSAT will provide simultaneous multi-wavelength observations of immense value in study of highly variable sources like X-ray binaries and the Active Galactic Nuclei. The review provides a brief summary of the payload characteristics of ASTROSAT and discusses some of the main science topics that will be addressed with particular emphasis on X-ray binaries and compact objects and the interface with gravitational wave experiments. The detection of GW is expected to strengthen multi-messenger astronomy. In another one of the reviews the aspects related to electromagnetic counterparts of GW sources are addressed.

A network of multi-kilometer length Interferometric GW Detectors have been built in the USA (LIGO) and Europe (Virgo, GEO600). These are Michelson interferometers with Fabry-Perot cavities in each arm of length 600 m for GEO600, 3 km for Virgo and 4 km for LIGO. When upgraded over the next five years to advanced detectors with more laser power, improved suspension system etc., the network will have the sensitivity to detect GWs from the coalescence of binary neutron stars at a distance of 200 Mpc and binary black holes at a redshift of z =0.4. Advanced detectors are expected to observe GW signals at monthly or even weekly rates. In the coming decade, an international network of multi-kilometer scale gravitational wave observatories with enhanced sensitivity will come into operation in the US, Europe, India and Japan. Working together, these ultra-sensitive detectors based on laser interferometry will begin to throw light on the astrophysics of gravitational wave sources, have implications for cosmology, probe the quantum limits to measurement and eventually unravel the mysteries of gravity as a fundamental force. The IndIGO consortium in close collaboration with LIGO Laboratory has been working to establish the third leg of LIGO in India: LIGO-India. The route towards LIGO-India and its current status is reviewed in another of the articles.

Beyond detection of GW and the astrophysical understanding of anticipated or even possibly exotic GW sources, GW observations will contribute to our understanding of the cosmology we live in and eventually to the fundamental nature of gravitation itself. How do we extend the successes of radio observations of binary pulsars to GW observations of compact binaries, EMRIs (Extreme Mass Ratio Inspirals) and MBH binaries in their inspiralling, coalescence and ringdown phases? The last two articles review comprehensively the progress made towards this exciting venture with emphasis on the testing of fundamental laws of spacetime. The volume rightfully concludes with these exciting possibilities that will be accessible over the next decades.

We thank all authors for having collaborated on this issue, reviewers for elaborate and punctual work, and especially Chee-Hok Lim and Roh-Suan Tung, the IJMPD editors, for their invaluable help in the success of the publication

# INTERNATIONAL JOURNAL OF MODERN PHYSICS D
*Vol. 22, No. 1 (January 2013)*

SPECIAL ISSUE
*Gravitational Wave Detection and Fundamental Physics in Space*
*Editors: Bala Iyer and Wei-Tou Ni*

TABLE OF CONTENTS